\documentclass[AMA,STIX1COL]{WileyNJD-v2}
\usepackage{dsfont}
\usepackage[utf8]{inputenc}

\usepackage{tikz}
\usepackage{adjustbox}
\usetikzlibrary{calc,trees,positioning,arrows,chains,shapes.geometric,%
    decorations.pathreplacing,decorations.pathmorphing,shapes,%
    matrix,shapes.symbols}

\tikzset{
>=stealth',
  punkt/.style={
    rectangle, 
    rounded corners, 
    fill=green!10,
    draw=black, very thick,
    text width=10em, 
    minimum height=3em, 
    text centered, 
    on chain},
  pil/.style={
    ->,
    thick,
    shorten <=2pt,
    shorten >=2pt,},
  line/.style={draw, thick, <-},
  element/.style={
    tape,
    top color=white,
    bottom color=blue!50!black!60!,
    minimum width=8em,
    draw=blue!40!black!90, very thick,
    text width=10em, 
    minimum height=3.5em, 
    text centered, 
    on chain},
  every join/.style={->, thick,shorten >=1pt},
  decoration={brace},
  tuborg/.style={decorate},
  tubnode/.style={midway, right=2pt},
}

\articletype{Commentary}%

\received{26 April 2016}
\revised{6 June 2016}
\accepted{6 June 2016}

\begin{document}

\title{Target estimands for population-adjusted indirect comparisons}

\author[1,2]{Antonio Remiro-Az\'ocar}

\authormark{REMIRO-AZ\'OCAR}

\address[1]{\orgdiv{Department of Statistical Science}, \orgname{University College London}, \orgaddress{\state{London}, \country{United Kingdom}}}

\address[2]{\orgdiv{Quantitative Research}, \orgname{Statistical Outcomes Research \& Analytics (SORA) Ltd}, \orgaddress{\state{London}, \country{United Kingdom}}}

\corres{*Antonio Remiro Az\'ocar, Department of Statistical Science, University College London, London, United Kingdom. \email{antonio.remiro.16@ucl.ac.uk}. Tel: (+44 20) 7679 1872. Fax: (+44 20) 3108 3105}

\presentaddress{Antonio Remiro Az\'ocar, Department of Statistical Science, University College London, Gower Street, London, WC1E 6BT, United Kingdom}

\abstract{Disagreement remains on what the target estimand should be for population-adjusted indirect treatment comparisons. This debate is of central importance for policy-makers and applied practitioners in health technology assessment. Misunderstandings are based on properties inherent to estimators, not estimands, and on generalizing conclusions based on linear regression to non-linear models. Estimators of marginal estimands need not be unadjusted and may be covariate-adjusted. The population-level interpretation of conditional estimates follows from collapsibility and does not necessarily hold for the underlying conditional estimands. For non-collapsible effect measures, neither conditional estimates nor estimands have a population-level interpretation. Estimators of marginal effects tend to be more precise and efficient than estimators of conditional effects where the measure of effect is non-collapsible. In any case, such comparisons are inconsequential for estimators targeting distinct estimands. Statistical efficiency should not drive the choice of the estimand. On the other hand, the estimand, selected on the basis of relevance to decision-making, should drive the choice of the most efficient estimator. Health technology assessment agencies make reimbursement decisions at the population level. Therefore, marginal estimands are required. Current pairwise population adjustment methods such as matching-adjusted indirect comparison are restricted to target marginal estimands that are specific to the comparator study sample. These may not be relevant for decision-making. Multilevel network meta-regression (ML-NMR) can potentially target marginal estimands in any population of interest. Such population could be characterized by decision-makers using increasingly available ``real-world'' data sources. Therefore, ML-NMR presents new directions and abundant opportunities for evidence synthesis.}


\maketitle

\renewcommand{\thefootnote}{\alph{footnote}}

\section{Introduction}

I recently participated in a very interesting discussion with Phillippo, Dias, Ades and Welton,\cite{remiro2021conflating, phillippo2021target} in response to their research article titled ``Assessing the performance of population adjustment methods for anchored indirect comparisons: A simulation study''.\cite{phillippo2020assessing} The original article presents an extensive simulation study evaluating the statistical performance of different population adjustment methods in health technology assessment (HTA). HTA requires comparing the effectiveness of all treatments that have been approved for a specific disease state, often in the absence of head-to-head randomized controlled trials (RCTs). In this context, ``indirect treatment comparisons'' are used to inform reimbursement decisions in many jurisdictions. 

The following situation is relatively common in HTA: (1) there are no head-to-head RCTs comparing the treatments of interest, but these treatments have been compared against a common comparator arm in separate RCTs; (2) patient-level data are unavailable for some of the studies. The indirect comparison is said to be ``anchored'' by the common comparator arm. In this scenario, the standard approaches\cite{bucher1997results, dias2013evidencedos} are biased where treatment effects are heterogeneous over variables (effect measure modifiers) that vary in distribution across studies, and the effects of these covariate differences do not balance each other out. Therefore, population adjustment methods are required to perform indirect comparisons that adjust for differences in covariate distributions across trials.\cite{phillippo2018methods} In the original simulation study,\cite{phillippo2020assessing} three methods are investigated: matching-adjusted indirect comparison (MAIC), simulated treatment comparison (STC), and a novel method recently proposed by the authors called multilevel network meta-regression (ML-NMR).\cite{phillippo2020multilevel}

In a recent editorial, co-authored with Anna Heath and Gianluca Baio,\cite{remiro2021conflating} I highlight that the different methodologies target distinct measures of effect. MAIC is based on propensity score weighting and targets a marginal treatment effect. STC and ML-NMR are outcome modeling-based methods, with effects estimated by the treatment coefficient of a multivariable regression. The typical implementation of STC targets a conditional treatment effect that, almost invariably, is incompatible in a pairwise indirect comparison, producing bias for non-collapsible measures of effect.\cite{remiro2021conflating,remiro2021methods} ML-NMR, developed by the authors of the simulation study,\cite{phillippo2020multilevel} extends outcome modeling to handle larger networks of treatments and studies. It targets a conditional treatment effect without the estimand compatibility issues of STC. In my original response,\cite{remiro2021conflating} I remark that the appropriateness of each methodology depends on the preferred inferential target, and that one should carefully consider whether a marginal or conditional treatment effect is of interest in a population-adjusted indirect comparison.

In their reply to my editorial, Phillippo et al. demonstrate that ML-NMR can also be used to target marginal treatment effects.\cite{phillippo2021target} Therefore, the method can support inference at the individual level and at the population level. This extension is a very relevant and impactful development for evidence synthesis, which will help overcome many limitations of pairwise indirect treatment comparisons in the estimation of marginal effects. Nevertheless, disagreement remains on what the target estimand for health technology assessment should be. In their response, Phillippo et al. comprehensively endorse the use of conditional treatment effect estimates to inform decision-making at the population level.\cite{phillippo2021target} However, I believe that estimates of the marginal treatment effect are necessary for population-level reimbursement decisions. Settling this debate is of central importance to offer a conclusion for policy-makers and applied practitioners in the field.  

\section{Target estimands in randomized controlled trials}\label{S2}

The objective of population-adjusted indirect comparisons (and, more generally, of evidence synthesis in HTA) is to emulate the analysis that would have been performed in an ideal head-to-head RCT, directly comparing the drugs of interest. By ``ideal'', I mean that the hypothetical RCT should have high internal validity, but also high external validity (this shall be discussed in Section \ref{S4}).\cite{deaton2018understanding, breskin2019using} This ideal RCT has been described as the ``target trial'' by Hern\'{a}n and Robins.\cite{hernan2016using} 

The RCT is widely considered the gold standard design for evaluating the efficacy of interventions due to its high internal validity,\cite{temple2000placebo} i.e., its potential for limiting bias within the study sample. Appropriate randomization guarantees covariate balance on expectation, so that the treatment groups are exchangeable and confounding is limited. Therefore, assuming no structural issues (e.g.~no dropout, informative missingness, measurement error, etc.), RCTs allow for unbiased estimation of the relative treatment effect within the sample.

There has been much debate over what the target estimand of an RCT should be. RCTs may target marginal or conditional estimands. These are calibrated at different hierarchical levels. The population-level marginal effect quantifies how mean outcomes change across all randomized individuals moving from one treatment to the other. On the other hand, conditional effects compare mean treatment outcomes across patients sharing similar covariate values. The marginal effect is often estimated by an ``unadjusted'' analysis. This may be a simple comparison of the expected outcomes for each group or a univariable regression including only the main treatment effect. RCTs typically report unadjusted analyses, which rely on measured and unmeasured covariates being balanced between treatment groups due to randomization.

The conditional treatment effect is often estimated as the treatment coefficient of an ``adjusted'' analysis, e.g.~a multivariable regression of outcome on the main effects of randomized treatment and a set of baseline covariates, such as prior medical history, demographic factors or physiological status. In this analysis, the target estimand is a weighted average of (also conditional) individual-level or subgroup-specific effects. The baseline covariates are pre-specified in the protocol or analysis plan and are likely to be prognostic variables, associated with the clinical outcome of interest. RCT analyses often adjust for one or more baseline covariates to correct for empirical confounding caused by chance imbalances between treatment groups. Incorporating this prognostic information can result in a more efficient use of data.\cite{senn1989covariate, hauck1998should, hernandez2004covariate, kahan2014risks} Note that, as highlighted by Daniel et al.,\cite{daniel2021making} ``the words conditional and adjusted (likewise marginal and unadjusted) should not be used interchangeably''. A recurring theme throughout this commentary is that marginal need not mean unadjusted.

An effect measure is collapsible when the marginal measure can be represented by a weighted average of subgroup-specific conditional measures.\cite{greenland2011adjustments,whittemore1978collapsibility,  greenland1999confounding,huitfeldt2019collapsibility} When effect measures are non-collapsible, there may be sizable differences between marginal and conditional estimands, even in a well-conducted RCT with covariate balance and the absence of confounding, no dropout, a very large sample size and no other structural issues.\cite{greenland2001confounding, greenland2011adjustments, kaufman2010marginalia} Non-collapsibility occurs in logistic regression analysis for the odds ratio,\cite{whittemore1978collapsibility,greenland1987interpretation, greenland1999confounding} in the Cox proportional hazards model for the hazard ratio,\cite{aalen2015does} and in other generalized linear models where the link function is neither the identity link (linear regression) nor the log link.\cite{gail1984biased}

Phillippo et al. use the following arguments to select the conditional estimand as the most appropriate inferential target for decision-makers in population-adjusted indirect comparisons:\cite{phillippo2021target}

\begin{enumerate}
    \item Conditional estimands account for differences in the distribution of prognostic covariates between groups but ``marginal estimands do not account for known population characteristics''.
    \item Conditional estimands have a ``population-average'' interpretation if treatment-by-covariate interactions are excluded from the analysis. 
    \item Conditional estimands are ``a more efficient choice'' than marginal estimands.
\end{enumerate}

I examine these points on a case-by-case basis. Point 1 conflates the terms ``marginal'' and ``unadjusted''. Estimates of the marginal effect need not be crude or unadjusted and may be covariate-adjusted.\cite{daniel2021making} Points 2 and 3 generalize conclusions based on covariate adjustment with linear regression, and do not apply to non-linear models with non-collapsible measures of effect. All of the points describe properties that are inherent to \textit{estimators} (the method of analysis), not \textit{estimands} (the target of the analysis).  

\subsection{Marginal is not synonymous with unadjusted}

Covariate-adjusted analyses have many advantages over unadjusted analyses. As stated by Phillippo et al.,\cite{phillippo2021target} the adjusted analysis is ``the recommended analysis that would be undertaken in the ideal (RCT) evidence scenario'' in the trials literature.\cite{senn1989covariate, hauck1998should, hernandez2004covariate, kahan2014risks} Nevertheless, the term ``marginal'' is not interchangeable with ``unadjusted''. Marginal is often interpreted as unadjusted and conditional as adjusted. However, the distinction between marginal and conditional describes the estimand, and that between adjusted and unadjusted relates to the estimator. It is true that unadjusted estimates of the marginal effect ignore any information on the distribution of prognostic covariates in the sample. Therefore, these cannot directly compensate for any lack of balance between treatment groups. However, estimates of the marginal effect can also be covariate-adjusted. In fact, covariate-adjusted marginal estimates are regularly used in the RCTs literature to correct for chance imbalances in baseline covariates and to improve precision.\cite{colantuoni2015leveraging, diaz2016enhanced, tsiatis2008covariate} 

Indeed, one can adjust for covariates using the outcome model and then average or ``standardize'' over a specific population to estimate marginal (but covariate-adjusted) effects that do compensate for lack of balance.\cite{daniel2021making, moore2011robust, jiang2019robust} For instance, in their reply, Phillippo et al.\cite{phillippo2021target} illustrate a simple method for ``marginalizing'' out the conditional effect estimates produced by ML-NMR where the outcome regression is a generalized linear model, by integration over the joint covariate distribution in the target population. The resulting covariate-adjusted estimate of the marginal effect would fully exploit the covariate information and account for imbalances in baseline characteristics, across studies and between treatment arms (in the trial with patient-level data). 

After all, ``population-adjusted'' indirect comparisons are ``covariate-adjusted'' indirect comparisons. MAIC uses weighting to produce a covariate-adjusted estimate of the marginal effect that is combined with a, typically unadjusted, marginal estimate in a pairwise indirect comparison. The covariate-adjusted marginal estimate accounts for covariate imbalances across studies but MAIC can be adapted to also account for imbalances between treatment arms.\cite{remiro2022twostage} Similarly, STC can incorporate G-computation or model-based standardization\cite{snowden2011implementation} so that a covariate-adjusted marginal estimate is produced that accounts for covariate imbalances across studies and treatment arms, and has no compatibility issues in the indirect treatment comparison.\cite{remiro2021parametric} As per the method proposed by Phillippo et al. for ML-NMR,\cite{phillippo2021target} one can estimate the marginal mean outcomes by treatment group, based on the fitted outcome regression. Then, the model-based predictions can be averaged over the joint covariate distribution of the target trial (that with aggregate-level data) and contrasted to produce an estimate of the marginal treatment effect that accounts for baseline covariates.\cite{remiro2021parametric} 

In summary, my original commentary does not call for crude unadjusted analyses over analyses adjusted for measured covariates.\cite{remiro2021conflating} Covariate-adjusted analyses may target marginal or conditional effects, and the debate is about covariate-adjusted estimates of the marginal estimand versus covariate-adjusted estimates of the conditional estimand.  

\textbf{On the preferences of regulatory agencies.} Phillippo et al. state that ``it is recommended practice to include prespecified prognostic factors in the analysis model''.\cite{phillippo2021target} Indeed, covariate adjustment is strongly advised by regulatory authorities such as the United States Food and Drug Administration (FDA)\cite{FDAguidance, FDAcovid} and the European Medicines Agency (EMA)\cite{EMAguidance} when approving new treatments. Nevertheless, this does not mean that the conditional estimand is the primary focus of such agencies. Recent FDA guidance seems to encourage the use of covariate-adjusted estimates of the marginal treatment effect as opposed to covariate-adjusted estimates of the conditional effect.\cite{FDAguidance} The addendum to the ICH E9 guidelines, Statistical Principles for Clinical Trials,\cite{ICHguidance} which introduces the estimands framework and has been adopted by the FDA and the EMA, discusses ``population-level summary measures'' of outcome as the primary target of inference in RCTs. This suggests a preference for marginal estimands (see subsection \ref{subsec22} of this commentary). 

\subsection{On the population-average interpretation of conditional estimands}\label{subsec22}

We shall define a ``population-average'' or ``population-level'' effect as one that compares mean outcomes in the entire trial population, i.e., between patients who have been randomly selected from this population, unconditional on their covariate values.\cite{hernan2020causal} Alternatively, an ``individual-level'' or ``subgroup-level'' effect is conditional on observed covariate values; it contrasts average outcomes within a stratum of patients that share the same covariates. 

Phillippo et al. argue that conditional estimands may have a ``population-average'' or an ``individual-level'' interpretation, depending on the method of analysis, and that I am conflating these effects.\cite{phillippo2021target} For instance, covariate adjustment through the ``analysis of covariance'' (ANCOVA) linear model specifies main effects but excludes treatment-by-covariate interactions. Therefore, the conditional treatment effect is assumed identical across covariate levels and is, therefore, not specific to subgroup membership. Conversely, with the inclusion of interaction terms, the conditional treatment effect is believed to differ across covariate values and the estimate may have an ``individual-level'' interpretation. It is argued that the ``population-average'' estimate of the conditional effect can be used to make population-level decisions.\cite{phillippo2021target}

I make two important remarks. Firstly, the homogeneity or uniformity of the conditional treatment effect estimate follows from statistical modeling assumptions, which may not be plausible. While the estimator may assume constancy across all patients, the true subgroup effects are not necessarily constant. If the constancy assumption does not hold, the ANCOVA conditional estimand no longer has a population-average interpretation, and corresponds to an ambiguous weighted average of subgroup-specific estimands.\cite{sloczynski2020interpreting} In any case, the preference for marginal or conditional estimands as inferential targets should not depend on the estimator. Conditional estimands can be well defined as subgroup-specific or individual-level effects, regardless of modeling assumptions. On the other hand, the population-level interpretation of the conditional estimate depends on such implicit assumptions. 

Secondly, the population-level interpretation of the conditional estimate in the case of no interaction relies on the measure of effect being collapsible. As mentioned earlier, an effect measure is collapsible when the marginal measure can be expressed as a weighted average of the subgroup-specific conditional measures. Mean differences in linear regression are collapsible across covariates. Without interaction terms, one assumes that there is no effect modification on the mean difference scale, such that mean differences are the same in every subgroup. In this case, the subgroup-specific mean difference estimates are also equal to the marginal mean difference estimate. The ``population-average'' interpretation of any conditional estimate simply reflects that it coincides with the marginal estimate due to collapsibility. Therefore, making a distinction\cite{phillippo2021target} between ``population-average'' conditional and ``population-average'' marginal estimands is not necessary. 

The situation is even more nuanced where the measure of effect is non-collapsible, as is the case for the (log) odds ratio in the original simulation study, which considers logistic regression models for binary outcomes.\cite{phillippo2020assessing} Unlike the mean difference, the (log) odds ratio is non-collapsible because the marginal measure cannot represent a weighted average of the individual- or subgroup-level conditional measures, even in the absence of confounding.\cite{greenland2001confounding, greenland1987interpretation, pang2016studying} Consider that a conditional (log) odds ratio estimate is derived from the treatment coefficient of a main effects logistic regression, assuming homogeneity. This estimate, which is equal to the constant subgroup-specific effect estimates, cannot have a population-level interpretation.\cite{greenland1987interpretation, gail1984biased, huitfeldt2019collapsibility, hauck1991consequence} This is a mathematical phenomenon that pertains to numeric properties of the measure of effect.\cite{greenland2011adjustments, greenland1999confounding} For the odds ratio, it is consequence of a special case of Jensen's inequality.\cite{pang2016studying} 


\textbf{The transportability of effect measures.} This is currently an area of debate. A general empirical observation is that conditional effects are more generalizable or transportable than marginal effects across different target populations,\cite{phillippo2021target} because marginal estimands change across different marginal covariate distributions. Nevertheless, conditional estimands also depend on the distribution of observed covariates if there is (conditional) effect modification on the scale of the effect measure. Both marginal and conditional estimands depend on the distribution of unmeasured covariates. 

In addition, another intuition is that conditional effects are less transportable because there may be many conditional estimands for a given population, one for every possible combination of baseline covariates considered for adjustment. Unless the conditional estimand is unambiguously defined prior to the analysis, it will be dependent on the specification of the selected adjustment model. Conditioning on different covariate sets leads to different conditional estimands, with their estimates not being comparable across studies.\cite{daniel2021making, martens2007conditioning} On the other hand, marginal estimands are typically defined without reference to a particular adjustment model. Pearl and Bareinboim\cite{pearl2014external} claim that marginal effects are more transportable than conditional effects, providing a mathematical proof. In the words of Daniel et al.,\cite{daniel2021making} this result highlights that ``any measured covariate can be adjusted for in the analysis, and then marginalized over according to any desired reference distribution, resulting in a marginal estimand that is just as transportable as any conditional estimand''. 

\subsection{Efficiency considerations}

The main argument by Phillippo et al. is that the conditional estimand is more efficient than the marginal estimand in the hypothetical evidence scenario described by the ideal RCT.\cite{phillippo2021target} It is claimed that the conditional is a more appropriate target estimand for health policy because it provides ``more efficient decision-making''.

It is true that for linear regression with maximum-likelihood estimation and continuous outcomes, a covariate-adjusted estimate of the conditional treatment effect should have a lower standard error than the unadjusted estimate of the marginal effect. The decrease in standard error is greater when the correlation between the baseline covariate(s) and outcome is strong, leading to a reduction in residual variance.\cite{assmann2000subgroup} Nevertheless, this is not the case when working with non-collapsible effect measures such as odds ratios in logistic regression with binary outcomes,\cite{gail1984biased,assmann2000subgroup, moore2009covariate, robinson1991some} or hazard ratios in Cox proportional hazards regression with survival outcomes.\cite{ford1995model,karrison2018restricted} These are two of the most widely used parameters, statistical models and outcome types in evidence synthesis in HTA.\cite{phillippo2019population} 

In these cases, adjusted estimates of the conditional estimand tend to have reduced precision and efficiency with respect to unadjusted estimates of the marginal estimand. For odds and hazard ratios, the covariate-adjusted maximum-likelihood estimator of a conditional effect has a standard error at least as large as the unadjusted maximum-likelihood estimator of a marginal estimand, in the ideal RCT.\cite{daniel2021making, robinson1991some, ford1995model} In fact, for non-collapsible effect measures, it is marginalized covariate-adjusted estimates that tend to have lower standard errors than both the original covariate-adjusted estimates of the conditional\cite{daniel2021making, moore2009covariate} and the unadjusted estimates of the marginal.\cite{colantuoni2015leveraging, diaz2016enhanced, tsiatis2008covariate} In addition, covariate-adjusted estimates of the marginal odds ratio seem to be less susceptible to small-sample and sparse-data bias than covariate-adjusted estimates of the conditional odds ratio.\cite{ross2021decreased} 

Pursuing greater precision and efficiency would make the covariate-adjusted marginal estimates more attractive than the original conditional estimates for non-collapsible effect measures in the ideal RCT. However, these comparisons are arguably inconsequential, because they are made for estimators targeting different estimands.\cite{daniel2021making, tsiatis2008covariate} One has to suitably define the target estimand before performing a ``like-for-like'' comparison of different estimators.\cite{lee2004conditional} When adjusted and unadjusted estimates of the marginal estimand are compared in the ideal RCT, covariate adjustment does increase precision by leveraging the prognostic information accounting for unexplained variation in the outcome. Adjustment for prognostic covariates does increase power to detect a non-null treatment effect with respect to an unadjusted analysis, regardless of whether a marginal or conditional estimand is targeted.\cite{hernandez2004covariate, assmann2000subgroup, robinson1991some} Power comparisons between marginal and conditional estimators are relevant because the corresponding estimands share the same null,\cite{daniel2021making, gail1988tests} e.g. under the null hypothesis of no treatment effect, the marginal odds ratio is one because the conditional odds ratio is one. 

A small but important aside: in contrast to the ``ideal RCT'' scenario, covariate adjustment increases the variance of indirect treatment comparisons.\cite{remiro2021methods} This is a desirable feature because standard unadjusted approaches\cite{bucher1997results} ignore \textit{cross-trial} differences in covariates that influence the outcome, thereby producing overly precise estimates and undercoverage.\cite{remiro2021methods, song2003validity} Covariate-adjusted indirect comparisons account for the additional uncertainty produced by these covariate differences. A reduction in precision is natural and necessary, and a function of the ``distance'' between the covariate distributions; we are trying to learn about a treatment effect in a different study than that in which it was originally investigated. Recent simulation studies show that standardized regression-adjusted (and in some cases, weighting-adjusted) estimates of the marginal (log) odds ratio are also more precise than regression-adjusted estimates of the conditional (log) odds ratio in this context.\cite{remiro2021parametric,remiro2020marginalization} These results are expected to hold for non-collapsible effect measures in general. 

I emphasize that the estimand of interest should be tailored to the scientific question that is being addressed. Namely, the choice of estimand should determine the estimator, and not vice versa. It makes sense to proceed sequentially, first determining the estimand that best answers the decision problem, and then using a method or analytic approach that is well suited for estimating it from the clinical trial data. Statistical efficiency should not drive the choice of the estimand. On the other hand, the estimand, unambiguously selected on the basis of relevance to decision-making, should drive the choice of the most statistically efficient estimator. This is because efficiency is a property inherent to estimators, not estimands.

\section{Target estimands in health technology assessment}\label{S3}

The development of pharmaceuticals is a multi-stage process, and HTA generally takes place late in this process. Section \ref{S2} has described factors that are relevant in preparation for the drug licensing stage. In this stage, the efficacy of a new medical technology is typically evaluated versus placebo or standard of care in an RCT. This trial may provide evidence supporting the regulatory approval of the drug by agencies such as the FDA and the EMA. In this setting, power considerations to test the ``no effect'' null hypothesis may also deserve attention. Indirect treatment comparisons are not typically applied for hypothesis testing or to obtain regulatory approval. These are highly underpowered in scenarios commonly met in practice\cite{weber2020comparison, kuhnast2017evaluation, mills2011estimating} and often lead to a conclusion of ``no clinical benefit''.\cite{ruof2014early} 

I now set aside the long-standing debate about target estimands in RCTs and focus on the decision problem at hand. Following regulatory approval, a pharmaceutical product can be submitted to HTA agencies worldwide, e.g.~the National Institute of Health and Care Excellence (NICE) in England and Wales, the Canadian Agency for Drugs and Technologies in Health, and the Pharmaceutical Benefits Advisory Committee in Australia. These formulate recommendations on whether health care technologies should be publicly funded by national health care systems. For HTA agencies, the demonstration of efficacy and/or effectiveness is necessary but not sufficient. Manufacturers must convince payers that their product offers the best ``value for money'' of all available options in the market.\cite{paul2001fourth} This requires comparing treatments in the absence of head-to-head RCTs.\cite{sutton2008use} The population adjustment methodologies evaluated in the original simulation study\cite{phillippo2020assessing} aim to quantify treatment effects in this scenario. The resulting estimates are used as inputs to health economic evaluations, e.g.~cost-effectiveness analyses comparing two or more competitor treatments. HTA bodies typically make reimbursement decisions on the basis of these evaluations. 

For instance, NICE operates an appraisal process in which companies submit evidence on both relative clinical and cost effectiveness. In this process, it is the mean cost and effectiveness at the population level that are relevant.\cite{thompson2000should} Evidence synthesis methods inform the mean treatment benefit in such analyses, where the main quantity of interest tends to be the incremental cost-effectiveness ratio (ICER), which is a population-level measure. For differential cost-effectiveness inferences and policies based on subject-level covariates, individual-level ICERs have been proposed.\cite{willan2001probability} However, these are currently of secondary interest in policy and may be problematic.\cite{o2002probability} Within centralized health care systems such as the National Health Service, planning entities and providers use the appraisal process to set population-level policies such as quality measures and clinical guidelines, and to select treatments for the population of patients within their remit. In this context, the marginal treatment effect is a more relevant target than the conditional effect. 

Conditional estimands would be of greater relevance in clinical practice or personalized/precision medicine, from the perspective of physicians making treatment decisions for individual patients. This is particularly the case if there is treatment effect heterogeneity and this is accounted for by the inclusion of treatment-by-covariate interactions. As advocated by Hauck et al.,\cite{hauck1998should} conditional treatment effects ``come as close as possible to the clinically most relevant subject-specific measure of effect''. For instance, physicians may be interested in how effective treatment is conditional on the age, gender and/or medical history of a particular patient, and may not desire to average over these characteristics. Indeed, marginal estimands make little sense in the context of clinical care and are not applicable in the context of decision-making for individual patients. 

If health care providers and reimbursement agencies were to make decisions at such level of granularity, conditional estimands would be of greater interest than marginal estimands. However, the research questions made by bodies such as NICE investigate how the average effect of an intervention impacts outcomes at the population level, and are used to make broad policy decisions and recommendations. In the health decision sciences, conditional treatment effects could also be of interest for individual-level microsimulation models that simulate the impact of interventions or policies on individual trajectories, which may be averaged out to estimate an overall population-level ICER. 

Finally, a very important consideration in health economic evaluation is uncertainty quantification,\cite{jackson2009accounting} where the assessment of parameter uncertainty is a central component.\cite{briggs2000handling, jackson2010structural} The conflation of marginal and conditional estimands is an issue for both collapsible and non-collapsible effect measures, because estimators targeting different estimands will produce different variance estimates. These variances quantify parameter uncertainty in cost-effectiveness analyses. Marginal and conditional estimates will quantify parameter uncertainty differently, and conflating these will lead to the incorrect propagation of uncertainty to the wider health economic decision model. This is particularly dangerous for probabilistic sensitivity analysis,\cite{griffin2011dangerous} a required component in the normative framework of HTA bodies such as NICE, used to characterize the impact of the model input uncertainty on decision-making.\cite{claxton2005probabilistic}

\section{External validity}\label{S4}

\subsection{Established population-adjusted indirect comparison methods}\label{S4S1}

In Section \ref{S3}, I established why marginal estimands should be preferred as inferential targets for population-level reimbursement decisions in HTA. In subsection \ref{subsec22}, I stated that there is only one well-defined marginal estimand for a specific population. Nevertheless, marginal estimands can change if the population definition is modified. As population-adjusted indirect comparisons assume treatment effect heterogeneity, different populations with different effect modifier distributions will have marginal (and conditional) estimands of distinct magnitudes. In addition, as I discuss in this section, most ``population-adjusted'' indirect comparisons refer to ``sample-adjusted'' indirect comparisons. This clouds the discussion of estimands further. 

MAIC and STC have been developed in the context of pairwise comparisons in a two-study scenario, where there is one ``index'' study with individual patient data (IPD) and another ``comparator'' study with unavailable IPD and only published aggregate-level data. A distinctive feature of MAIC and STC is that, due to patient-level data limitations, the methods contrast treatments in the comparator study sample, defined by the summary moments of baseline characteristics in ``Table 1'' of the publication. Inferences can only be interpreted within this sample-specific context, which imposes constraints on the marginal estimand that is targeted. 

As currently conceptualized, MAIC and STC imply that the comparator study sample on which inferences are made is exactly the study's target population. Alternatively, the assumption is that the study sample is a random sample, i.e., representative, of such population, ignoring sampling variability in the patients' baseline characteristics and assuming that no random error attributable to such exists. In reality, the subjects of the comparator study have been sampled from a, typically more diverse, target population of eligible patients, defined by the trial's inclusion and exclusion criteria. 

Random sampling is seldom feasible in recruitment strategies for trial participants. For instance, individuals with health-seeking behaviors are more likely to enroll in the trial. Candidates who meet the trial eligibility criteria may not be invited to participate.\cite{dahabreh2016using} Conversely, invited study-eligible individuals may not provide informed consent, an ethical necessity for enrollment, and choose not to participate. In summary, it is rarely the case that a study sample is a random sample of the target population of the study, because trial participation is subject to convenience sampling, and volunteerism or self-selection.\cite{dahabreh2016using, weiss2019generalizing} This is a problem of external validity with respect to the target population of the trial. 

A more important limitation is that, even if the comparator study sample is representative of the study's target population, such target population may be systematically different to the target population of policy interest,\cite{weiss2019generalizing, tipton2013improving} i.e., the group of patients who will receive the intervention in routine clinical practice. The populations targeted by clinical trials tend to be more narrowly defined and less heterogeneous in composition and health status,\footnote{An exception are so-called ``pragmatic'', ``effectiveness'' or ``practical'' trials.\cite{van2014opportunities, choudhry2017randomized} These are large-scale multi-center trials carried out in ``real-world'' settings. Their patients tend to be more representative of decision-making target populations because the trials have broad eligibility criteria and design elements that promote enrollment of a wide range of participants. However, this may come at the expense of lower adherence and higher drop-out rates.} to maximize power in efficacy and safety testing, and to enhance statistical precision and efficiency.\cite{rothwell2005external,rothwell2010commentary,greenhouse2008generalizing} In addition, the comparator study may have been conducted in many separate geographical regions, different to that of relevance for HTA decision-making. Payers reimburse treatments at the local market level and HTA decisions are likely to concern local patient populations.\cite{happich2020reweighting} These are questions of external validity with respect to the target population for the decision.

\begin{figure}[!htb]
\begin{adjustbox}{width=18cm,center}
\begin{tikzpicture}
  [node distance=.9cm,
  start chain=going below,auto]
  \node[punkt, fill=red!10] (market) {Index study sample};
  \node[punkt, inner sep=5pt,below=0.9cm of market, fill=red!10] (formidler) {Index study population};
  \node[punkt,right=0.9cm of market, fill=green!10] (market2){Comparator study sample};
  \node[punkt, inner sep=5pt,below=0.9cm of market2, fill=green!10] (formidler2) {Comparator study population};
  \node[punkt,right=0.9cm of market2, fill=blue!10] (market3){``Real-world'' sample};
  \node[punkt, inner sep=5pt,below=0.9cm of market3, fill=blue!10] (formidler3) {Decision-making target population};
  \node[above=of market] (dummy) {};
 \draw[<->,
      >          = latex,
      line width = .8pt] (-1.5,1)--++
                         (12.4,0)node[midway,
                                                  above]{Transportability};
 \draw[<-,dashed,
      >          = latex,
      line width = .8pt] (-2.5,0.05)--++
                         (0,-2.55)node[midway,
                                                   left]{Sampling}; 
 \draw[->,
      >          = latex,
      line width = .8pt] (11.9,0.05)--++
                         (0,-2.55)node[midway,
                                                   right]{Generalizability};
  \end{tikzpicture}
\end{adjustbox}
\caption{External validity addresses whether inferences can be extended beyond specific samples. Researchers make a distinction between generalizability and transportability.\cite{degtiar2021review} Generalizability entails generalizing the findings from an RCT to the population from which the trial participants were drawn, i.e., the RCT sample is a proper subset of the trial-eligible population.\cite{tipton2013improving, cole2010generalizing, stuart2011use} Transportability involves translating inferences to an external target sample or population.\cite{pearl2014external, rudolph2017robust, westreich2017transportability, petersen2011compound} This
diagram has been inspired by Degtiar and Rose.\cite{degtiar2021review}
}\label{fig1}
\end{figure}
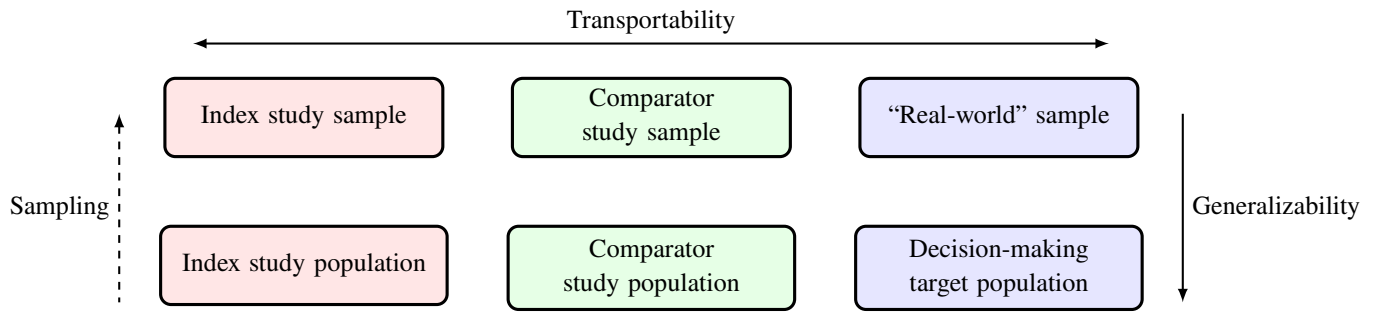

In terms of estimands, let's consider three distinct marginal estimands. MAIC and (a marginalized version of) STC would target a sample-average marginal estimand in the comparator study. However, this may not coincide with the population-average marginal estimand for the target population of the study. In turn, this may not match the population-average marginal estimand for the relevant target population required for HTA decision-making. If the samples or populations are non-exchangeable, an internally valid estimate for the marginal estimand in one sample/population is not necessarily unbiased for the marginal estimand in the others.\cite{westreich2019target, imai2008misunderstandings} The relationships between the different samples and populations are displayed in Figure \ref{fig1}. Moving horizontally, pairwise methods such as MAIC are limited to transporting inferences from the index study sample to the comparator study sample. 

\subsection{ML-NMR: new directions for evidence synthesis?}

ML-NMR presents abundant opportunities for evidence synthesis. Following the ``marginalization'' extension by Phillippo et al.,\cite{phillippo2021target} it allows for the estimation of marginal estimands in any of the study samples included in the meta-analysis. For instance, one can produce estimates in the most heterogeneous study, which may be more representative of the target population for HTA decision-making. Alternatively, one can produce estimates in the most homogeneous study to avoid overlap violations. Better yet, inferences are not necessarily restricted to one of the studies included in the meta-regression. The target sample or population could be generated from an external data source. Presuming that this contains the covariates that have been adjusted for in the studies, inferences can be transported to the target of interest. Crucial but demanding assumptions are that the outcome meta-regression is correctly specified and, consequently, that there is conditional exchangeability across settings. As with all methods, sensitivity analyses exploring alternative model specifications to the base-case are essential.   

The target could be defined by HTA policy-makers for the specific disease under study. It could be characterized by the joint covariate distribution observed in an observational sample or in secondary health data sources such as disease registries, cohort studies and insurance claims databases. Such administrative datasets have high cross-sectional richness, and are typically larger, less selected, and more representative of target populations of interest than the participants recruited into trials.\cite{corrigan2018real, sturmer2020methodological, girman2019considerations} Electronic health records from hospital systems are also valuable tools to define appropriate ``real-world'' targets.\cite{ramsey2020using} These are compelling due to several aspects, including reasonably large sample sizes and a high degree of clinical detail, specificity and breadth.\cite{carrigan2020using, chau2020developing} HTA bodies such as NICE are increasingly using ``real-world data'' to identify representative populations and inform or update assessments of effectiveness and cost-effectiveness.\cite{jaksa2021organized} These covariate data are likely to be available at the time of the HTA appraisal process. 

The sponsor of the index study, submitting evidence to HTA bodies, could use pairwise methods such as MAIC to weight or standardize its results to the external target source provided by decision-makers. However, IPD are unavailable for the comparator trial, for both the manufacturer submitting the evidence and the HTA agency assessing the evidence. Hence, the submitting company cannot weight or standardize the comparator study results to the external target. Therefore, it is challenging for pairwise methods like MAIC and STC, as currently conceptualized, to facilitate an indirect comparison in this target. 

Finally, we should ask ourselves to what substantive population are indirect treatment comparisons supposed to apply to. The main premise of population-adjusted indirect comparisons is to provide equipoise, removing bias due to covariate differences across studies and comparing effect estimates in the same population. However, because treatment effects are assumed heterogeneous, any discussion of equipoise needs to be framed within the question: ``equipoise to whom?''. Equipoise in a particular study sample does not guarantee equipoise in the target population for decision-making. Any claim about equipoise is ill-formed without reference to the desired target population for inference. The same applies to any claim about representativeness. Unfortunately, most applications of population-adjusted indirect comparisons do not explicitly describe the target population of the analysis, as found by a recent review of HTA appraisals.\cite{phillippo2019population} 

\section{Conclusions}

Ultimately, both marginal and conditional estimands have their advantages and disadvantages, and their validity is situation-specific, dependent on the type of inference desired. Both population-level and individual-level inferences are complementary to each other in order to fully understand how treatments compare. Researchers should articulate their question of interest, choose the estimand that best answers this question, and select an appropriate methodology on the basis of such choice. In HTA submissions, marginal estimands should be targeted because agencies make reimbursement decisions at the population level. 

Nevertheless, the marginal estimands targeted by pairwise methods such as MAIC are specific to the comparator study sample. ML-NMR targets marginal estimands more flexibly, and can extend within-study inferences to the relevant target population for HTA decision-making. In addition, outcome modeling-based methods such as ML-NMR are typically more precise and efficient than weighting methods in estimating marginal treatment effects, particularly where overlap is poor and effective sample sizes after weighting are small.\cite{phillippo2021target,remiro2021parametric,robins1992estimating} Another advantage of outcome modeling approaches is that they can estimate conditional and marginal estimands, as the conditional estimates can be standardized into marginals. On the other hand, weighting-based methods such as MAIC are restricted to marginal inference. 

Finally, ML-NMR is applicable in treatment networks of any size with the aforementioned two-study scenario as a special case. This is important because a recent review finds that 56\% of NICE technology appraisals include larger networks,\cite{phillippo2019population} where the standard pairwise population-adjusted indirect comparisons cannot be readily applied. For all these reasons, ML-NMR presents novel and exciting opportunities for evidence synthesis in HTA.

\section*{Acknowledgments}

The author thanks the editor and reviewers of Statistics in Medicine for their insightful comments, which helped improved the article. The author is hugely thankful to his PhD supervisors, Gianluca Baio and Anna Heath, who provided valuable feedback, and his PhD examiners, Nick Latimer and Manuel Gomes. The examiners' remarks motivated the need for these clarifications. 

\subsection*{Financial disclosure}

No funding to report. 

\subsection*{Conflict of interest}

The author declares no potential conflict of interests.

\subsection*{Data Availability Statement}

Data sharing is not applicable to this article because no new data were analyzed.


\bibliography{wileyNJD-AMA}

\end{document}